\documentclass[preprint,showpacs,preprintnumbers,amsmath,amssymb]{revtex4}
\usepackage{graphicx}
\usepackage{dcolumn}
\usepackage{bm}
\usepackage{feynmf}

\begin{document}
\preprint{MTA-PHYS-0504 \cr UWO-TH/05-05}


%
%

\title{ISOSPIN SYMMETRY BREAKING IN $B->K^*\gamma$ DECAY DUE TO AN EXTRA GENERATION OF VECTOR QUARKS}

\author{MOHAMMAD R. AHMADY}

\address{Department of Physics, Mount Allison University, 67 York Street\\
Sackville, New Brunswick, Canada E4L 1E6\\
mahmady@mta.ca}

\author{FARRUKH CHISHTIE}

\address{Department of Applied Mathematics, University of Western Ontario\\
London, Ontario, Canada N6A 5B7\\
fchishti@uwo.ca}


\begin{abstract}
The extra contributions due to an extra generation of vector-like quarks to the isospin symmetry breaking observable associated with the radiative $B\to
K^*\gamma$ decay is obtained.  It is shown that this additional contribution is sensitive to the nonunitarity parameter $U^{sb}$, which is a measure of the the
strength of the non-zero tree-level flavor changing neutral current in this model.  The significance of this result is that, once accurate experimental results
on the isospin asymmetry becomes available, one can constrain the $U^{sb}$ independent of the mass of the extra quarks and so a much more stringent acceptable
model parameter space could be obtained.

\end{abstract}

\pacs{13.20.He, 12.39.St, 12.15.Mm}

\keywords{isospin asymmetry; vector-like quarks; tree-level FCNC.}

\maketitle

\section{Introduction}

The Standard Model (SM) of electroweak interactions had been quite successful in describing the experimental data up to recent past.  However, there are now some
indications of possible inconsistencies between this model and experiment.  In order to improve our understanding of what lies beyond the SM, we need to make
progress on two fronts: 1) improve the accuracy of the experimental data 2) reduce the uncertainties in the theoretical predictions of the physical observables
based on the SM.  The hope is to find the deviations from the SM on as many front as possible so that to be able to rule in or out the various scenarios for the
new physics.

One of the main sources of theoretical error in making predictions involving hadrons is the nonperturbative nature of the strong interactions. Since the weak
interactions involving quarks is one of the main venues to test the SM, it is quite essential to be able to deal with this issue. Due to confinement, the
physical states that appear in the matrix element of the quark weak currents are hadrons, which are bound states of quarks, rather than free quarks themselves.
The form factors that parameterize these matrix elements characterize our lack of knowledge of the strong force in the nonperturbative regime and can not be
evaluated analytically from first principles.  However, with the help of the fast computers, lattice QCD has been able to make significant progress in numerical
calculation of some of the nonperturbative parameters. Nonetheless, to reduce the theoretical uncertainly, one solution for the time being is to look for
observables which are not too sensitive to form factors and other nonperturbative parameters.

The other criteria for an interesting process for testing the SM is the sensitivity to new physics.  Usually, one looks for a process which receives vanishingly
small contribution from the SM and therefore, any significant deviation from zero in the experimental measurement could be attributed to the new physics beyond
the SM.  On the other hand, experimental confirmation of the null result to a good precision should rule out a good number of new physics scenarios if the
process satisfies the above criteria.

The radiative inclusive decay $B\to X_s\gamma$, which is due to the underlying flavor changing neutral current (FCNC) quark transition $b\to s\gamma$, and its
exclusive mode $B\to K^*\gamma$ have proven to be quite important processes for examining the SM and new physics since they were first observed by CLEO. The most
recent experimental data on these transitions are as follows\cite{belle,babar,cleo}:
\begin{eqnarray}
Br(B\to K^{*0}\gamma)=\displaystyle\left\{ \begin{matrix}{(4.01\pm 0.21\pm 0.17 )\times 10^{-5} \;\;\; (BELLE)}\\ {(3.92\pm 0.20\pm 0.24)\times 10^{-5}\;\;\;
(BABAR)}\\ {(4.55 ^ {+0.72}_{-0.68} \pm 0.34)\times 10^{-5} \;\;\; (CLEO)} \end{matrix} \right . \nonumber \\ Br(B^+\to K^{*+}\gamma)=\displaystyle\left\{
\begin{matrix}{(4.25\pm 0.31\pm 0.24 )\times 10^{-5} \;\;\; (BELLE)}\\ {(3.87\pm 0.28\pm 0.26)\times 10^{-5}\;\;\; (BABAR)}\\ {(3.76 ^ {+0.89}_{-0.83} \pm
0.28)\times 10^{-5} \;\;\; (CLEO)} \end{matrix} \right.
\end{eqnarray}
The current PDG average for the inclusive mode from CLEO and the BELLE measurements is\cite{pdg}
\begin{equation}
Br(B\to X_s\gamma)=(3.3\pm 0.4)\times 10^{-4}
\end{equation}

\begin{figure}[pb]
\includegraphics{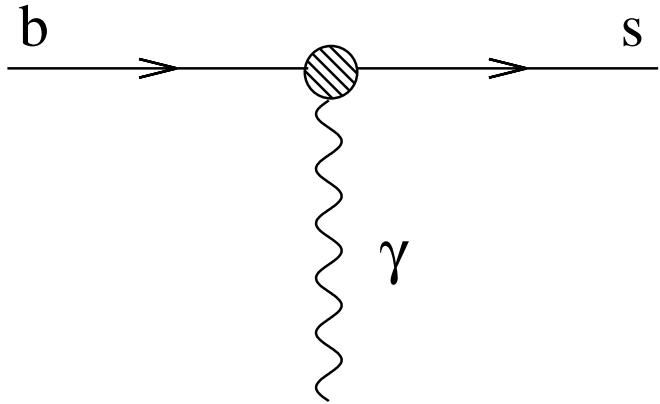} \vspace*{8pt} \caption{The effective $b\to s\gamma$ transition. \label{f1}}
\end{figure}

\begin{figure}[pb]
\includegraphics{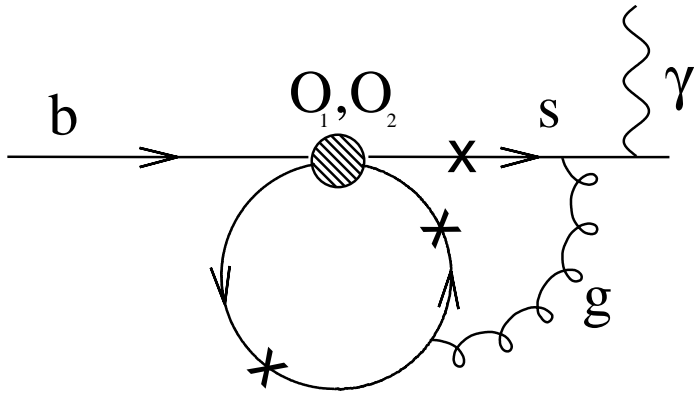} \vspace*{8pt} \caption{${\cal O} (\alpha_s)$ current-current operators contribution to $b\to s \gamma$. Cross
marks are the alternative locations for the coupling of the emitted photon. \label{f2}}
\end{figure}

\begin{figure}[pb]
\includegraphics{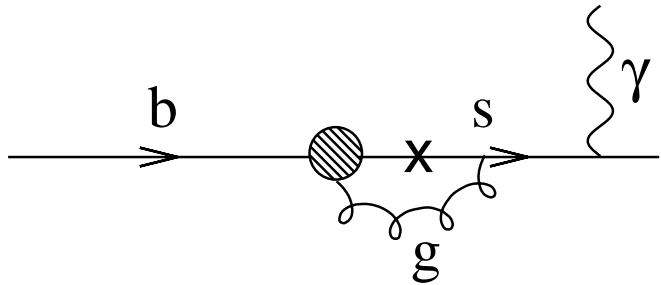} \vspace*{8pt} \caption{${\cal O} (\alpha_s)$ chromomagnetic penguin operator contribution to $b\to s \gamma$.
Cross mark is the alternative location for the coupling of the emitted photon.\label{f3}}
\end{figure}

This radiative transition is dominated by the one-loop quantum effect in SM which in effective Hamiltonian language is represented by the magnetic dipole
operator as illustrated in fig. \ref{f1}.
Since the exotic particles that appear in various extensions of the SM can appear as quantum fluctuations in the loop, this rare B decay has been at the focus of
many theoretical investigations to constrain different new physics scenarios.  The inclusive decay, though easier to handle theoretically, is much more difficult
to be accurately measured.  To leading order in $\alpha_s$, one can equate the $B\to X_s\gamma$ equal to the underlying quark transition $b\to s\gamma$ (fig.
\ref{f1}).  The systematic approach to decays at energies much below the $W$-boson mass, like B-meson decays, is through the effective Hamiltonian.  Here the
heavy fields, like $W$ and top-quark fields, are integrated out of the electroweak lagrangian and the effective Hamiltonian is written in terms of a series of
operators with increasing mass dimensions.  Of course, practically, only the lowest and next-to-lowest mass dimension operators contribute to the transitions.
These operators are listed as follows:
\begin{itemize}
\item{Current-current operators:
\begin{eqnarray}
\nonumber O_{1u}&=&\bar q_\alpha\gamma^\mu Lu_\alpha\bar u_\beta\gamma_\mu Lb_\beta \; ,\;
O_{2u}=\bar q_\alpha\gamma^\mu Lu_\beta\bar u_\beta\gamma_\mu Lb_\alpha \;\; ,\\
\nonumber O_{1c}&=&\bar q_\alpha\gamma^\mu Lc_\alpha\bar c_\beta\gamma_\mu Lb_\beta \; ,\; O_{2c}=\bar q_\alpha\gamma^\mu Lc_\beta\bar c_\beta\gamma_\mu
Lb_\alpha \;\; .
\end{eqnarray}}
\item{QCD penguin operators:
\begin{eqnarray}
 O_3&=&\bar q_\alpha\gamma^\mu Lb_\alpha\sum_{q'}\bar q_\beta'\gamma_\mu Lq_\beta' \; ,\; O_4=\bar q_\alpha\gamma^\mu
Lb_\beta\sum_{q'}\bar q_\beta'\gamma_\mu Lq_\alpha' \;\; ,
\\ O_5&=&\bar q_\alpha\gamma^\mu Lb_\alpha\sum_{q'}\bar q_\beta'\gamma_\mu Rq_\beta '\; ,\;
O_6=\bar q_\alpha\gamma^\mu Lb_\beta\sum_{q'}\bar q_\beta'\gamma_\mu Rq_\alpha' \;\; .
\end{eqnarray}}
\item{Electroweak penguin operators:
\begin{eqnarray}
\nonumber O_7&=&\frac{3}{2}\bar q_\alpha\gamma^\mu Lb_\alpha\sum_{q'=u,d,s,c,b}e_{q'}\bar q_\beta'\gamma_\mu Rq_\beta'\;\; ,\\
\nonumber O_8&=&\frac{3}{2}\bar q_\alpha\gamma^\mu Lb_\beta\sum_{q'=u,d,s,c,b}e_{q'}\bar q_\beta'\gamma_\mu Rq_\alpha' \;\; ,\\
O_9&=&\frac{3}{2}\bar q_\alpha\gamma^\mu Lb_\alpha\sum_{q'=u,d,s,c,b}e_{q'}\bar q_\beta'\gamma_\mu Lq_\beta '\;\; ,\\ \nonumber O_{10}&=&\frac{3}{2}\bar
q_\alpha\gamma^\mu Lb_\beta\sum_{q'=u,d,s,c,b}e_{q'}\bar q_\beta'\gamma_\mu Lq_\alpha' \;\;.
\end{eqnarray}}
\item{Electro- and chromo-magnetic operators:
\begin{eqnarray}
O_\gamma &=&\frac{e}{4\pi^2}\bar q_\alpha\sigma^{\mu\nu}(m_bL+m_sR)b_\alpha F_{\mu\nu} \;\; , \\
O_g &=&\frac{g_s}{4\pi^2}\bar q_\alpha\sigma^{\mu\nu}(m_bL+m_sR)T^a_{\alpha\beta}b_\beta G_{\mu\nu}^a \;\; .
\end{eqnarray}
}
\end{itemize}
In the above equations $q=s\; {\rm or}\; d$ and $L(R)=\frac{1-(+)\gamma_5}{2}$ are the projection operators.  As mentioned above, $O_\gamma$ has the main
contribution to the $b\to s\gamma$ transition in the leading order.  At the next-to-leading order in $\alpha_s$, besides the electromagnetic penguin operator,
the current-current operators $O_1$ and $O_2$ (fig. \ref{f2}) and the chromomagnetic penguin operator $O_g$(fig. \ref{f3}) contribute as well.
This leads to the SM prediction of $Br(B\to X_s\gamma )=(3.6\pm 0.30)\times 10^{-4}$ and consequently provides constraints on the new physics beyond the
SM.\cite{gm}.

The exclusive $B\to K^*\gamma$ decay is more difficult to calculate theoretically due to the nonperturbative nature of the hadronic matrix element of the above
operators.  These matrix elements are parameterized in terms of a number of form factors.  One could use certain approximate symmetries of the QCD, like heavy
quark symmetry, to reduce the number of these parameters.  But in any case, the calculation of these parameters introduces model dependence and theoretical
uncertainty in our predictions.

\begin{figure}[pb]
\includegraphics{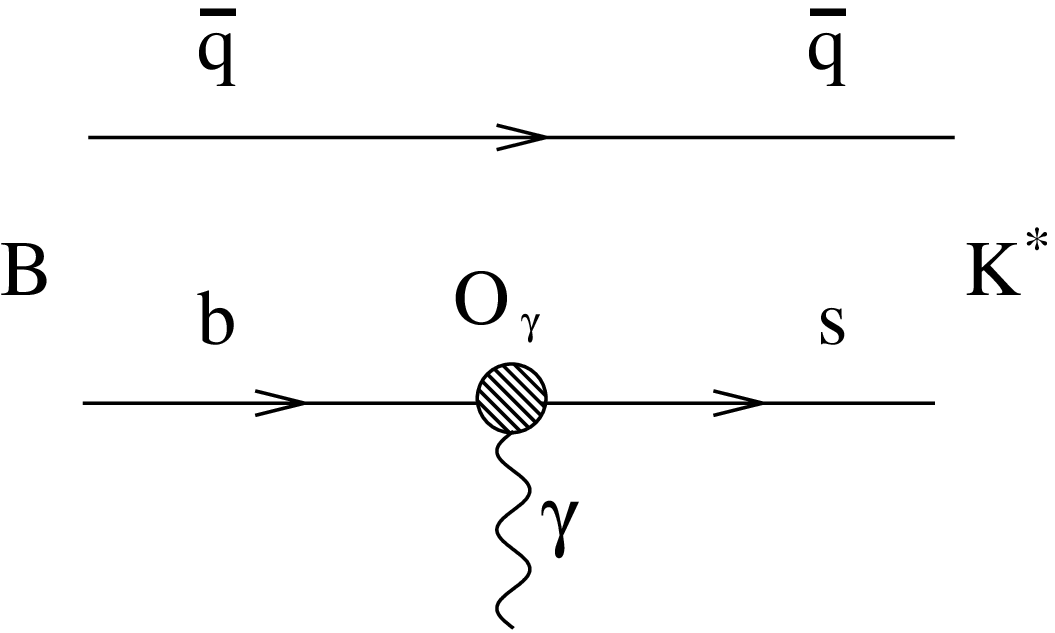} \vspace*{8pt} \caption{Leading order factorizable contribution to $B\to K^* \gamma$.\label{f4}}
\end{figure}

\begin{figure}[pb]
\includegraphics{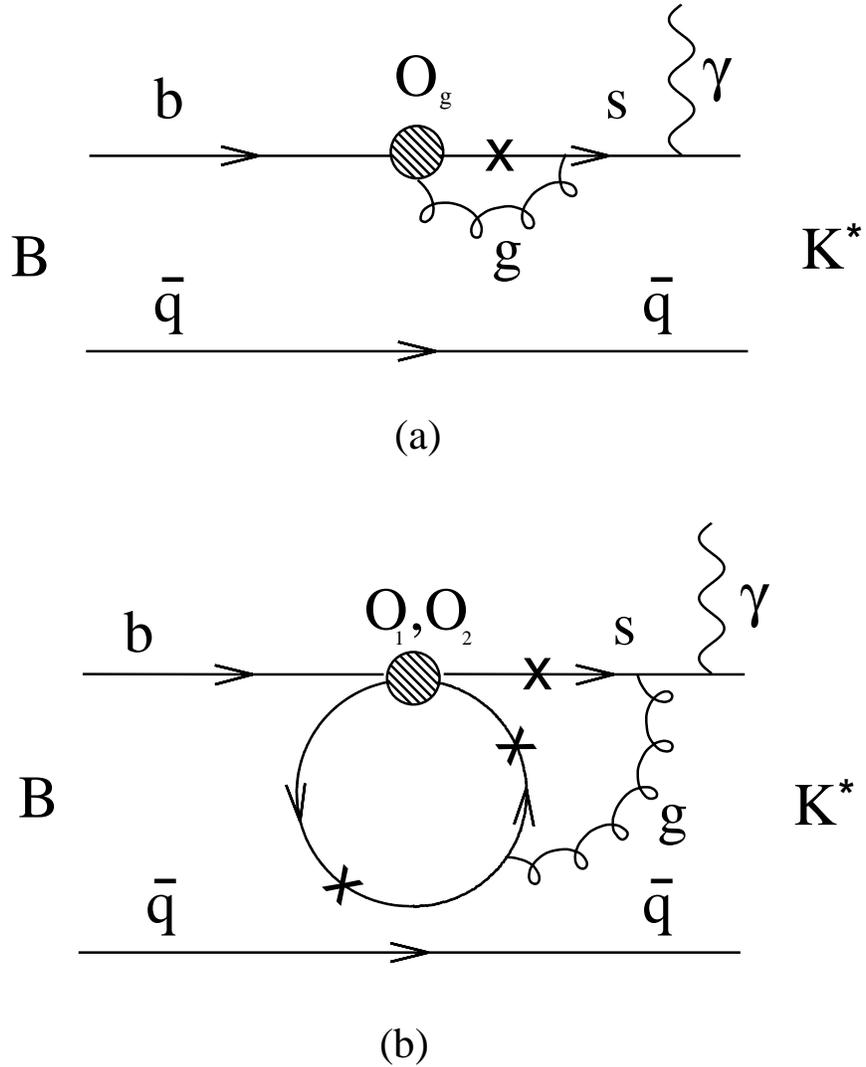} \vspace*{8pt} \caption{Next-to-leading order factorizable contributions to $B\to K^* \gamma$. Cross marks are the
alternative locations for the coupling of the emitted photon.\label{f5}}
\end{figure}

\begin{figure}[pb]
\includegraphics{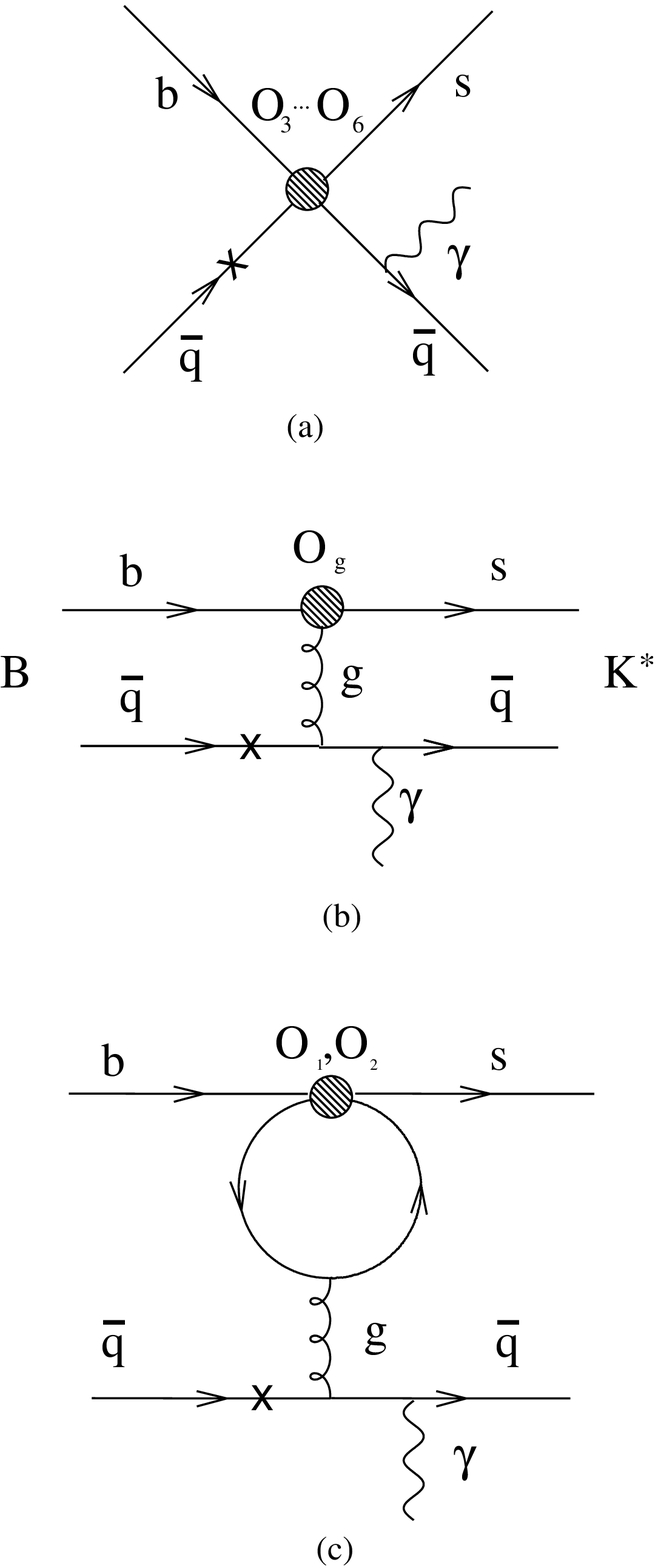} \vspace*{8pt} \caption{Nonfactorizable contributions to $B\to K^* \gamma$. Cross mark is the alternative location
for the coupling of the emitted photon.\label{f6}}
\end{figure}

\begin{figure}[pb]
\includegraphics{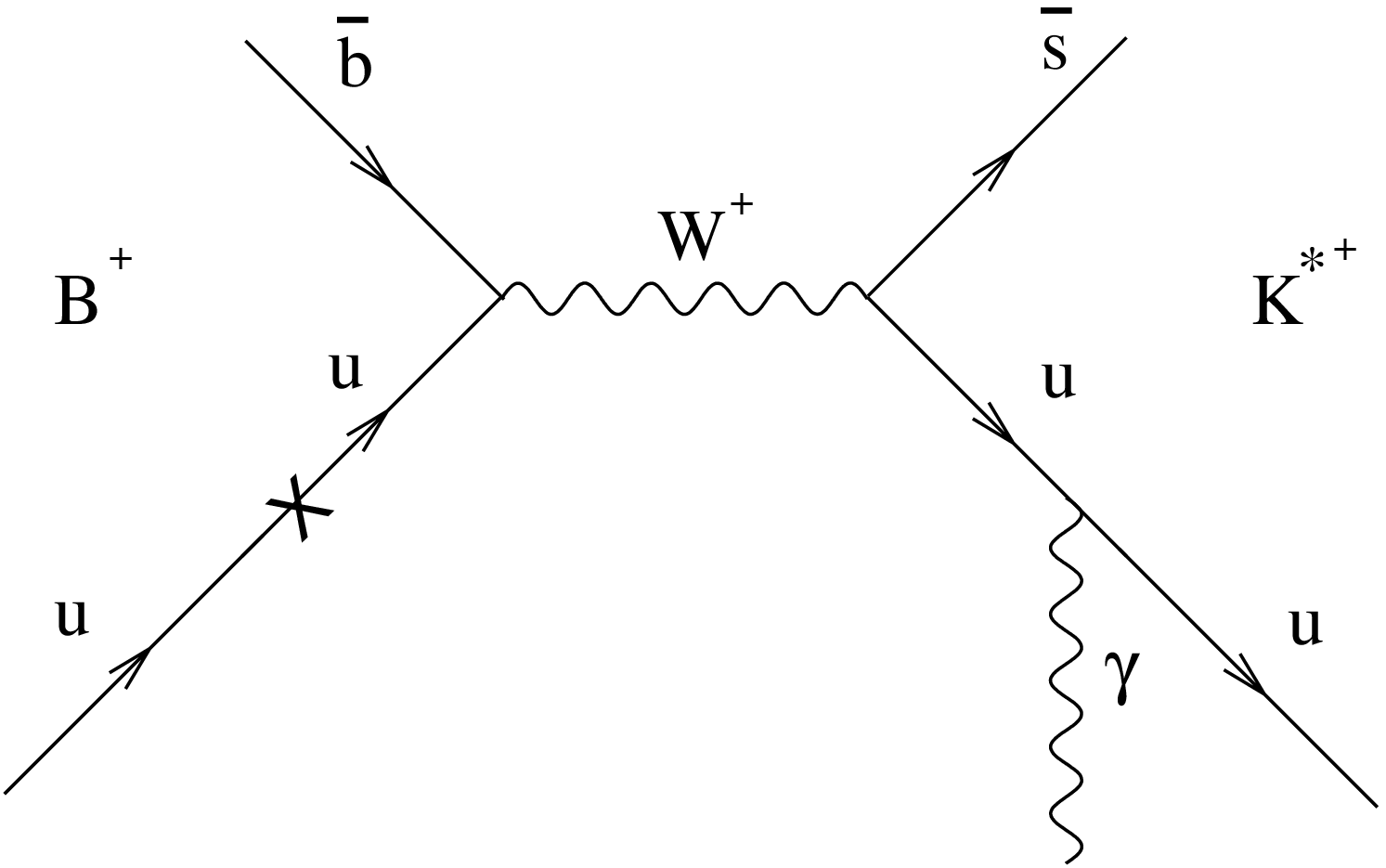} \vspace*{8pt} \caption{Annihilation diagram contributing to the $B^+\to K^{*+}\gamma$ decay.\label{f7}}
\end{figure}

In Ref. \cite{bb}, a model independent approach based on QCD factorization in the leading power of $\Lambda_{QCD}/m_B$ has been proposed.  Here, the operators
that were mentioned previously in connection to the inclusive decay contribute to the factorizable term to the leading (fig. \ref{f4}) and next-to-leading order
in $\alpha_s$ (fig. \ref{f5}). On the other hand, there are new operators involving the light quark in the B meson that contribute to the nonfactorizable term as
illustrated in fig. \ref{f6}).  Also, there are annihilation diagrams contributing to the nonfactorizable decay of the charged B mesons (fig. \ref{f7}), however
these transitions are CKM as well as power ($\Lambda_{QCD}/m_B$) suppressed.
The nonperturbative QCD enters through a number of universal functions, a form factor for $B\to K^*$ transition and light-cone distribution amplitudes for $B$
and $K^*$ mesons.  In fact, due to the sensitivity of the branching ratio to the flavor of the light quark through the nonfactorizable terms in the ones depicted
in fig. \ref{f6}, one obtains a slightly different branching ratio for neutral and charged B mesons. Taking the appropriate value for the input parameters,
authors in Ref. \cite{bb} obtain $Br(B^-\to K^{*-}\gamma )=7.45\times 10^{-5}$ as compared to $Br(\bar B^0\to \bar K^{*0}\gamma )=7.09\times 10^{-5}$.

\section{Isospin Symmetry Breaking in $B\to K^*\gamma$}
Besides the branching ratio, one could associate another observable, the isospin symmetry breaking, to the radiative $B\to K^*\gamma$ decay.  This asymmetry is
define as follows:
\begin{equation}
\Delta_{0-}=\frac{\Gamma (\bar B^0\to\bar K^{*0}\gamma ) -\Gamma (B^-\to K^{*-}\gamma )}{\Gamma (\bar B^0\to\bar K^{*0}\gamma )+\Gamma (B^-\to K^{*-}\gamma
)}\;\; , \label{isospinasym}
\end{equation}
with $\Delta_{0+}$ obtained from eq.(\ref{isospinasym}) by using the charge conjugate modes.  $\Delta_{0\pm}$ could prove to be important observables for
examining the SM as well as discriminating between various new physics scenarios\cite{petrov}. The data from Belle\cite{belle} and Babar\cite{babar} point to
isospin asymmetries of at most a few percent and consistent with zero within the experimental error:
\begin{eqnarray}
\Delta_{0-}&=& +0.051\pm 0.044({\rm stat.})\;\pm 0.023({\rm sys.})\;\pm 0.024(R^{+/0})\;\;\; (Babar)\; , \label{babar}\\
\Delta_{0+}&=& +0.012\pm 0.044({\rm stat.})\;\pm 0.026({\rm sys.})\;\;\; (Belle)\; , \label{belle}
\end{eqnarray}
 where the last error in eq.(\ref{babar}) is due to the uncertainty in the ratio of the branching fractions of the neutral and charged B meson production
in $\Upsilon (4S)$ decays. As we noticed in the previous section, this asymmetry in the SM is due to the non-spectator contributions and has been estimated
within the QCD factorization approach in Refs. \cite{bb} and \cite{kn} and the perturbative QCD method in Ref. \cite{kms}. In general, to the lowest order in
$\alpha_s$ and the small parameter $\Lambda_{QCD}/m_B$, current-current, QCD penguin and chromomagnetic penguin operators make the dominant contributions to the
isospin symmetry breaking within the SM.  The role of the electroweak penguin operators are more or less the same as their QCD counterparts, however, since their
corresponding Wilson coefficients are significantly smaller, they can be safely ignored. The more accurate measurement of the isospin asymmetry in the near
future and a better understanding of the SM prediction for this observable should provide a sensitive testing venue for possible models of new physics. One such
model is the extension of the SM with an extra generation of iso-singlet quarks\cite{ahmadynagashima,ac} which is explained in the next section.

\section{Vector-like quark model(VQM)}
In this model, the gauge structure of the SM remains intact except for an additional pair of iso-singlet quarks, which we denote them by $U$ and $D$.  The
difference between these new quarks and ordinary quarks of the three SM generations is that, unlike the latter ones, both left- and right-handed components of
the former quarks are $SU(2)_L$ singlets.  Therefore, the Dirac mass terms of vector-like quarks, i.e.
\begin{equation}
m_U(\bar U_L U_R+\bar U_R U_L)+m_D(\bar D_L D_R+\bar D_R D_L) \;\; ,
\end{equation}
are invariant under electroweak gauge symmetry.  However, the masses of the ordinary quarks arise from their gauge invariant Yukawa couplings to an iso-doublet
scalar Higgs field $\phi$ as follows:
\begin{equation}
-f^{ij}_d{\bar\psi_L}^id_R^j\phi -f^{ij}_u{\bar\psi_L}^iu_R^j\tilde\phi \; +\; H.C. \;\; ,
\end{equation}
where $i,j=1,2,3$ is covering the three generations of the regular quarks, and the doublet of fermions $\psi_L^i$ is defined as
\begin{equation}
\psi_L^i\equiv {\left (\begin{matrix}{
                       u^i} \cr
                       {d^i} \end{matrix}\right )}_L \;\; .
\end{equation}
At the same time, additional $SU(2)_L$ invariant Yukawa couplings between vector-like and ordinary quarks, in the form
\begin{equation}
-f^{i4}_d{\bar\psi_L}^iD_R\phi -f^{i4}_u{\bar\psi_L}^iU_R\tilde\phi \; +\; H.C. \;\; ,
\end{equation}
leads to mixing among 4 up- and down-type quarks of the same charge.  As a result, after spontaneous electroweak symmetry breaking due to $\langle \phi \rangle
=v\neq 0$, we obtain the following mass terms:
\begin{equation}
{\bar d_L}^\alpha M_d^{\alpha\beta}d_R^\beta + {\bar u_L}^\alpha M_u^{\alpha\beta}u_R^\beta \; +\; H.C. \;\; ,
\end{equation}
where $M_d$ and $M_u$ being $4\times 4$ mass matrices and $\alpha ,\beta =1..4$ cover ordinary and vector-like quarks.  In general, the mass matrices are not
diagonal and unitary transformations from weak to mass eigenstates are necessary to achieve diagonalization.  Denoting the mass eigenstates with $u_{L,R}'$ and
$d_{L,R}'$, we have
\begin{equation}
u^\alpha_{L,R}={A^u_{L,R}}^{\alpha \beta}{u'_{L,R}}^\beta \; ,\;  d^\alpha_{L,R}={A^d_{L,R}}^{\alpha \beta}{d'_{L,R}}^\beta \;\; ,\label{trans}
\end{equation}
where the unitary transformation matrices $A^{u,d}_{L,R}$ are chosen such that ${A^d_L}^\dagger M_d A_R^d$ and ${A^u_L}^\dagger M_u A_R^u$ are diagonal.  The
interesting property of the VQM is that the transformations (6) lead to inter-generational mixing among quarks not only in the charged current sector but also in
the neutral current interactions.  This is due to the fact that the extra iso-singlet quarks carry zero weak isospin and thus, are not involved in $SU(2)_L$
interactions as weak eigenstates.  For example, the charge current interaction term
\begin{equation}
{J^W_{CC}}^\mu=\sum^3_{i=1}I\frac{g}{\sqrt{2}}{\bar u_L}^i\gamma^\mu d_L^iW^+_\mu \; +\; H.C. \;\; ,
\end{equation}
transforms to
\begin{equation}
{J^W_{CC}}^\mu=\sum_{\alpha ,\beta =1}^4 I\frac{g}{\sqrt{2}}{\bar {u'}_L}^\alpha V^{\alpha \beta}\gamma^\mu {d'_L}^\beta W^+_\mu \; +\; H.C. \;\; ,
\end{equation}
where
\begin{equation}
V^{\alpha \beta}=\sum_{i=1}^3{({A^u_L}^\dagger)}^{\alpha i}{(A_L^d)}^{i\beta}\;\; ,\label{V}
\end{equation}
when expressed in terms of mass eigenstates via eq. (\ref{trans}).  $V$ is the $4\times 4$ generalization of the Cabibbo-Kobayashi-Maskawa(CKM)\cite{ckm} quark
mixing matrix.  The fact that the fourth generation is iso-singlet, i.e. $i=1,2,3$, leads to non-unitarity of the mixing matrix $V$ as demonstrated in the
following:
\begin{eqnarray}
\nonumber {(V^\dagger V)}^{\alpha \beta}&=&\sum_{\delta =1}^4 {V^{\delta\alpha}}^*V^{\delta\beta}=\sum_{\delta =1}^4\sum_{i,j=1}^3{\left [{({A^u_L}^{i\delta})}^*
{A^d_L}^{i\alpha}\right ]}^*{({A^u_L}^{j\delta})}^*{A^d_L}^{j\beta} \\
\nonumber &=&\sum_{i,j=1}^3{({A^d_L}^{i\alpha})}^*{A^d_L}^{j\beta}\sum_{\delta =1}^4{({A^u_L}^{j\delta})}^* {A^u_L}^{i\delta} \\
&=&\sum_{i=1}^3{({A^d_L}^{i\alpha})}^*{A^d_L}^{i\beta}=\delta^{\alpha\beta}-{({A_L^d}^{4\alpha})}^*{A_L^d}^{4\beta}\;\; .\label{unitary1}
\end{eqnarray}
In obtaining the last line in (\ref{unitary1}), the unitarity of the tranformation matrix $A_L^u$ has been utilized. In the same way, one can show
\begin{equation}
{(VV^\dagger )}^{\alpha\beta}=\delta^{\alpha\beta}-{({A_L^u}^{4\alpha})}^*{A_L^u}^{4\beta}\;\; .\label{unitary2}
\end{equation}
Equations (\ref{unitary1}) and (\ref{unitary2}), together with the unitarity of $A^{u,d}_L$, indicate that the quark mixing matrix $V$ {\it can not} be unitary.
This property of the VQM has interesting consequences in the neutral current sector where non-vanishing tree level FCNC, proportional to the deviation of the
quark mixing matrix (eq. (\ref{V})) from unitarity, are generated.  To demonstrate this explicitly, let us examine the neutral current which is coupled to
$Z_\mu$ boson
\begin{equation}
{J^Z_{NC}}^\mu=I\frac{g}{\cos\theta_w}\left (I^q_w\sum_{i=1}^3{\bar q_L}^i\gamma^\mu q^i_L-Q_q\sin^2\theta_w\sum_{\delta=1}^4({\bar q_L}^\delta\gamma^\mu
q_L^\delta +{\bar q_R}^\delta\gamma^\mu q_R^\delta )\right ) \;\; ,\label{ncurr}
\end{equation}
where $Q_q$ is the electric charge of the quark $q$.  The first term in (\ref{ncurr}) is proportional to $I^q_w$, the third component of the isospin, which has
the value $+1/2$ or $-1/2$ for the up- or down-type quarks, respectively.  Consequently, the iso-singlet quarks, which have zero isospin, {\it do not} contribute
to this term.  As a result, under the transformations (\ref{trans}), the neutral current (\ref{ncurr}) can be expressed in terms of the mass eigenstates as
follows:
\begin{eqnarray}
\nonumber {J^Z_{NC}}^\mu&=&I\frac{g}{\cos\theta_w}\left (I^q_w\sum_{i=1}^3\sum_{\alpha ,\beta =1}^4{\bar{q'}_L}^\alpha\gamma^\mu {q'_L}^\beta
{({A_L^q}^\dagger )}^{\alpha i}{A_L^q}^{i\beta} \right. \\
\nonumber &{}&\left. -Q_q\sin^2\theta_w\sum_{\delta =1}^4\sum_{\alpha ,\beta =1}^4 {\bar{q'}_L}^\alpha\gamma^\mu {q'_L}^\beta {({A_L^q}^\dagger )}^{\alpha
\delta}
{A_L^q}^{\delta\beta}+{\bar{q'}_R}^\alpha\gamma^\mu {q'_R}^\beta {({A_R^q}^\dagger )}^{\alpha \delta}{A_R^q}^{\delta\beta} \right ) \\
&=&I\frac{g}{\cos\theta_w}\sum_{\alpha ,\beta =1}^4\left (I^q_wU^{\alpha\beta}{\bar{q'}_L}^\alpha\gamma^\mu {q'_L}^\beta
-Q_q\sin^2\theta_w\delta^{\alpha\beta}{\bar{q'}}^\alpha\gamma^\mu {q'}^\beta \right ) \;\; ,\label{ncurr2}
\end{eqnarray}
where
\begin{eqnarray}
\nonumber U^{\alpha\beta}=\sum_{i=1}^3{({A_L^q}^{i\alpha})}^*{A_L^q}^{i\beta}&=&\delta^{\alpha\beta}-{({A_L^q}^{4\alpha})}^*{A_L^q}^{4\beta} \\ &=&\left
\{\begin{matrix}{
                       {(V^\dagger V)}^{\alpha\beta}\; ,\; q\equiv {\rm down-type}} \cr
                       {{(V V^\dagger)}^{\alpha\beta}\; ,\; q\equiv {\rm up-type}}\end{matrix} \right. \;\; .\label{U}
\end{eqnarray}
We observe that the non-unitarity of the mixing matrix $V$ in the VQM leads to the tree level FCNC in the $Z$ sector.  This in turn results in additional
contributions to the isospin symmetry breaking.  In fact, in the following section, we show that the isospin asymmetry in $B\to K^*\gamma$ transitions offers an
excellent physical observable for constraining the parameters of the VQM.  As is shown in our result, the advantage here is that this asymmetry, unlike the
branching ratio for $B\to K^*\gamma$ for example, is sensitive to only one model parameter, namely the non-unitarity parameter $U^{sb}$. Therefore, isospin
asymmetry can provide a good constraint on the size of the FCNC in the context of VQM irrespective of the masses of the additional quarks.

\section{Isospin asymmetry in $B\to k^*\gamma$ within the VQM}
\begin{figure}
\includegraphics{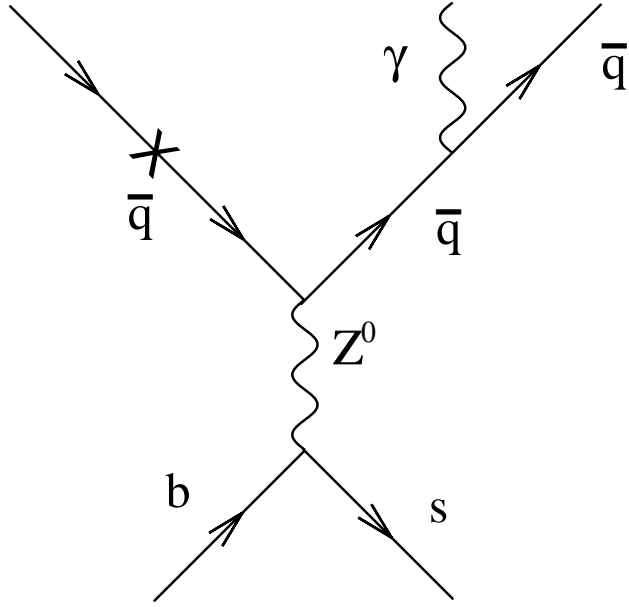} \vspace*{8pt} \caption{Tree level contribution to non-spectator process in $B\to K^* \gamma$.  Cross represents
the alternative coupling of the emitted photon. \label{f8}}
\end{figure}

The non-vanishing FCNC at the tree level leads to an additional contributing Feynmann diagram of fig. \ref{f8}.  The amplitude for $b\bar q\to s\bar q$ via $z^0$
exchange in VQM can be written as\cite{ac}:
\begin{eqnarray}
A^{VQM}&=&\frac{ig}{2\cos (\theta )}\left(-\frac{1}{2}U^{sb}\right)\bar s\gamma^\mu (1-\gamma_5)b\times\frac{1}{M_Z^2} \nonumber \\
&&\frac{ig}{2\cos (\theta )}\left[ (I^q_W-Q_q{\sin^2{\theta }})\bar q\gamma_\mu (1-\gamma_5)q-Q_q{\sin^2{\theta}}\bar q\gamma_\mu (1+\gamma_5)q\right]\;\;
,\label{vqmamp}
\end{eqnarray}
where $U^{sb}=(V^\dagger V)^{sb}$ is a measure of the non-unitarity of the extended quark mixing matrix as derived in eq. \ref{U}. One can then write
(\ref{vqmamp}) in terms of the effective operators $O_3$ and $O_5$ and therefore the presence of the extra vector quarks is summarized in some additional terms
in the Wilson coefficients $C_3$ and $C_5$ to the leading order in the strong coupling $\alpha_s$.
\begin{eqnarray}
C_3^{VQM}&=&\frac{U^{sb}}{V_{tb}V_{ts}^*}(I_W^q-Q_q\sin^2{\theta})\nonumber \\  &=&\frac{U^{sb}}{V_{tb}V_{ts}^*}\displaystyle\left\{^{\displaystyle
1/2-2/3\sin^2{\theta}
=0.35\ldots q=up}_{\displaystyle -1/2+1/3\sin^2{\theta}=-0.42\ldots q=down}\right. \;\; ,\nonumber \\
C_5^{VQM}&=&-\frac{U^{sb}}{V_{tb}V_{ts}^*}Q_q\sin^2{\theta}=\frac{U^{sb}}{V_{tb}V_{ts}^*}\displaystyle\left\{^{\displaystyle -2/3\sin^2{\theta}=-0.15\ldots
q=up}_{\displaystyle 1/3\sin^2{\theta}=0.08\ldots q=down}\right. \;\; . \label{wilsoncoef}
\end{eqnarray}
With the upper bound $\vert U^{sb}\vert\lesssim 10^{-3}$ coming from the rare B decays \cite{ahmadynagashima}, the additional contribution due to the tree level
FCNC could be comparable to the SM value of these coefficients at $\mu=m_b$, i.e. $C_3=0.014$ and $C_5=-0.041$.

Here an explanation is in order.  Strictly speaking, one should include these extra terms, which are proportional to the electric charge of the light quark, in
the electroweak penguin operators $O_{7\ldots 10}$\cite{hiller}.  However, since, as far as isospin symmetry breaking to the leading order of $\alpha_s$ is
concerned, one can ignore these operators within SM, we prefer to write the additional VQM-generated contributions in terms of the dominant QCD penguin
operators.  In any case, our result does not change had we followed the strict formulation of the problem.

Following the method of Ref. \cite{kn}, the nonspectator isospin symmetry breaking contribution can be written as $A_q=b_qA_{lead}$, with $q=u\;{\rm or}\; d$
being the flavor of the light anti-quark in the B meson.  $A_{lead}$ is the leading isospin symmetry conserving spectator amplitude. As mentioned in the
introduction, to the leading order in the strong coupling constant $\alpha_s$, the main contribution to $B\to K^*\gamma$ is from the electromagnetic penguin
operator $O_\gamma$ and the factorizable amplitude $A_{lead}$ is proportional to the form factor $T_1^{B\to K^*}$ which parameterizes the hadronic matrix element
of this operator to the leading order in $\Lambda_{QCD}/m_b$. $b_q$ is the parameter that depends on the flavor of the spectator.  In fact, this parametrization
of the nonspectator (nonfactorizable) contribution leads to a simple expression of the isospin asymmetry of eq. (\ref{isospinasym}) in terms of $b_q$:
\begin{equation}
\Delta_{0-}=\Re (b_d-b_u)\;\; ,\label{asymb}
\end{equation}
Using the expression for $b_q$ which is derived in Ref \cite{kn} within the QCD factorization method, we obtain the share of vector quarks to the isospin
asymmetry as follows:
\begin{equation}
\Delta_{0-}^{VQM}=\Re\left(\frac{4\pi^2f_B}{m_bT_1^{B\to
K^*}a_7^c}\frac{U^{sb}}{V_{tb}V_{ts}^*}\frac{1}{N_c}\left[0.38\frac{f_{K^*}^{\bot}F_{\bot}}{m_b}-0.28\frac{f_{K^*}m_{K^*}}{6\lambda_Bm_B}\right]\right)\;\;
.\label{asymvqm}
\end{equation}
The numerical input for the parameters of eqn. (\ref{asymvqm}) are tabulated in Table \ref{table1}, which results in an isospin asymmetry due to the extra
generation of quarks of the form:
\begin{equation}
\Delta_{0-}^{VQM}=-0.07\Re\left(\frac{U^{sb}}{V_{tb}V_{ts}^*}\right)\;\; . \label{asymnum}
\end{equation}
The significance of eq. (\ref{asymnum}) is that it is sensitive only to one model parameter, i.e. $U^{sb}$, and therefore, with more precise experimental data
becoming available in the future, this observable could serve to impose a stringent constrain on the all important nonunitarity parameter of the vector quark
model.

\begin{table}
\caption{The numerical values of the parameters in eq. (\ref{asymvqm}).} {\begin{tabular}{@{}ccccccccc@{}} \toprule
$T_1$&$m_b$&$\lambda_B$&$f_{K^*}$&$f_{K^*}^{\bot}$&$m_B$&$m_{K^*}$&$F_{\bot}$&$a_7^c$ \\
\colrule
0.32& 4.2 GeV & 0.35 GeV&0.226 GeV&0.175 GeV&5.28 GeV&0.892 GeV&1.21&-0.41 \cite{bb}\\
\botrule
\end{tabular} \label{table1}}
\end{table}

\section{Conclusion}
Isospin asymmetry in $B\to K^*\gamma$ decay mode could prove to be an important observable in the precision test of the SM and constraining various scenarios for
new physics beyond it.  Since this asymmetry is defined as the ratio of branching fractions, its sensitivity to the nonperturbative parameters could be minimal.
In this work, we calculated the additional contribution to the isospin asymmetry due to an extra generation of vector quarks and derived an expression in terms
of the model parameter $U^{sb}$ within the QCD factorization method.  A precise measurement of $BR(B\to K^*\gamma)$ for both neutral and charged B-mesons should
impose a stringent constraint of this important VQM model parameter

\section*{Acknowledgments}

M.A. would like to thank Dr. Amir Fariborz and SUNYIT for the excellent organization of this year's MRST conference and two summer research students at Mount
Allison, Ben Angus and Casey Steele for their help in preparing this manuscript.  F.C. would like to thank the hospitality of Subh-e-Nau Environmental Research
Center where part of this research was carried out.  M.A.'s research is partially funded by a discovery grant from NSERC.



\begin{thebibliography}{0}    

\bibitem{belle}  BELLE Collab. (M. Nakao, {\it et al}.), {\it Phys. Rev.} {\bf D69}, 112001 (2004).

\bibitem{babar}  BABAR Collab. (E. Paoloni) , talk at the Rencontres de Moriond, Electroweak Session, La Thuile, Italy, March 2004, hep-ex/0406083.

\bibitem{cleo}   CLEO Collab. (T.E. Coan, {\it et al}.), {\it Phys. Rev. Lett.} {\bf 84}, 5283 (2000).

\bibitem{pdg}   Particle Data Group (S. Eidelman, {\it et al}.), {\it Phys. Lett.} {\bf B592}, 1 (2004).

\bibitem{gm}  Paolo Gambino and Mikolaj Misiak, {\it Nucl. Phys.} {\bf B611}, 338 (2001).

\bibitem{bb} Stefan W. Bosch and Gerhard Buchalla, {\it Nucl. Phys.} {\bf B621}, 459 (2002).

\bibitem{petrov} Alexey A. Petrov, {\it Phys. Lett.} {\bf B399}, 172 (1997); John F. Donoghue, Alexey A. Petrov, {\it Phys. Rev.} {\bf D53}, 3664 (1996).

\bibitem{kn} A. Kagan and M. Neubert, {\it Phys. Lett.} {\bf B539}, 227 (2002).

\bibitem{kms} Y.Y. Keum, M. Matsumori and A.I. Sanda, {\it hep-ph/0406055}.

\bibitem{ahmadynagashima} M. Ahmady, M. Nagashima and A. Sugamoto, {\it Phys. Rev.} {\bf D64}, 054011 (2001).

\bibitem{ac} M. Ahmady and F. Chishtie, {\it hep-ph/0507114}.

\bibitem{ckm} M. Kobayashi and T. Maskawa, {\it Prog. Theor. Phys.} {\bf 49}, 652 (1973); \\
N. Cabibbo, {\it Phys. Rev. Lett.} {\bf 10}, 531 (1963).

\bibitem{hiller} David Atwood and Gudrun Hiller, {\it hep-ph/0307251}.

\end{thebibliography}
\end{document}